\documentclass[aps,
prb,
reprint,
floatfix,
]{revtex4-2}

\usepackage{graphicx}
\usepackage{physics}
\usepackage{booktabs}
\usepackage{amsthm,amssymb,mathrsfs}
\usepackage{tensor}
\usepackage{bm}
\usepackage{dsfont}
\usepackage[caption=false]{subfig}

\usepackage{hyperref} 
\hypersetup{colorlinks=true,allcolors=blue}

\usepackage{xargs} 
\usepackage{xcolor}

\newcommand{\iu}{\mathrm{i}} 
\newcommand{\e}{\mathrm{e}} 
\newcommand{\D}{\mathrm{d}} 
\newcommand{\hc}{\mathrm{h.c.}} 
\newcommand{\sign}{\operatorname{sgn}} 

\let\v\temp %

\let\v\relax
\newcommand{\v}[1]{\ensuremath{\mathbf{#1}}} 

\newcommand{\normord}[1]{\ensuremath{{\bm{:}}\mathbin{#1}{\bm{:}}}}

\makeatletter
\newcommand*{\coloneqq}{\mathrel{\rlap{%
			\raisebox{0.28ex}{$\m@th\cdot$}}%
		\raisebox{-0.28ex}{$\m@th\cdot$}}%
	=}
\newcommand*{\eqqcolon}{=\mathrel{\rlap{%
			\raisebox{0.28ex}{$\m@th\cdot$}}%
		\raisebox{-0.28ex}{$\m@th\cdot$}}%
}
\makeatother

\makeatletter
\newsavebox{\@brx}
\newcommand{\llangle}[1][]{\savebox{\@brx}{\(\m@th{#1\langle}\)}%
	\mathopen{\copy\@brx\mkern2mu\kern-0.9\wd\@brx\usebox{\@brx}}}
\newcommand{\rrangle}[1][]{\savebox{\@brx}{\(\m@th{#1\rangle}\)}%
	\mathclose{\copy\@brx\mkern2mu\kern-0.9\wd\@brx\usebox{\@brx}}}
\makeatother

\makeatletter
\newcommand*{\rom}[1]{\expandafter\@slowromancap\romannumeral #1@}
\makeatother

\begin{document}
	\title{
        Fate of odd-parity magnetism in one dimension
    }

    \author{Kasper Rettedal Eikeland} 
	\affiliation{Center for Quantum Spintronics, Department of Physics, Norwegian University of Science and Technology, NO-7491 Trondheim, Norway}
    
	\author{Sondre Duna Lundemo} 
	\affiliation{Center for Quantum Spintronics, Department of Physics, Norwegian University of Science and Technology, NO-7491 Trondheim, Norway}
	
	\author{Asle Sudb\o}
	\email[Corresponding author: ]{asle.sudbo@ntnu.no}
	\affiliation{Center for Quantum Spintronics, Department of Physics, Norwegian University of Science and Technology, NO-7491 Trondheim, Norway}
	
	\date{\today} 
	
	\begin{abstract}
        We consider a one-dimensional model for a $p$-wave magnet within the bosonization framework.
        The model consists of itinerant electrons described by an extended Hubbard model coupled to a chain of localized moments through the Kondo exchange. 
        The classical ground state of the local-moment system captures the salient features of an odd-parity magnet.
        By bosonizing the coupled system, a description in terms of coupled Luttinger liquids follows, giving rise to a rich weak-coupling phase diagram. 
        It is shown that the spin chain develops quasi long-range order consistent with the combined time-reversal and translation symmetry defining the $p$-wave magnet.
        We highlight the peculiar role played by this order in establishing a commensurability condition on the electronic filling under which additional interactions appear in the bosonic field theory.
        It is demonstrated that these interactions endow the electron spectral function with a pronounced $p$-wave character.
        Away from commensurate filling, these interactions are rendered irrelevant and the ensuing $p$-wave character is lost. 
	\end{abstract}
	
	\maketitle 
	
	\section{Introduction}\label{sec:intro}

    An odd-parity magnet refers to an itinerant magnet in which the electron bands are spin-split with an odd parity, for instance $p$-wave  \cite{Brekke-Linder-MinimalModelsTransport-2024,Hellenes-Smejkal-PwaveMagnets-2024}.
    These systems have recently attracted much attention for their potential use in spintronics applications \cite{Song-Comin-ElectricalSwitchingPwave-2025,Yamada-Hirschberger-MetallicPwaveMagnet-2025,Dsouza-Christensen-OddparityMagnetismFebased-2026,Sukhachov-Linder-Coexistence$p$waveMagnetism-2025,Hodt-Linder-Fate$p$waveSpin-2025}, following the discovery of the collinear, even-parity spin-split counterpart known as altermagnets.
    While breaking time-reversal symmetry, an altermagnet preserves the combination of time reversal with a spatial lattice rotation or inversion \cite{Smejkal-Sinova-CrystalTimereversalSymmetry-2020,Smejkal-Jungwirth-EmergingResearchLandscape-2022,Smejkal-Jungwirth-ConventionalFerromagnetismAntiferromagnetism-2022}.
	This property lifts Kramers degeneracy and yields electron bands with spin splitting consistent with the rotational symmetry \cite{Krempasky-Jungwirth-AltermagneticLiftingKramers-2024,Reimers-Jourdan-DirectObservationAltermagnetic-2024,Lee-Kim-BrokenKramersDegeneracy-2024}.
	In contrast, the $p$-wave character of the spin-polarized bands in $p$-wave magnets is protected by a combination of time-reversal symmetry $\mathcal{T}$ with a lattice translation $\tau$ by half a unit cell \cite{Brekke-Linder-MinimalModelsTransport-2024, Zeng-Wang-OddparityAltermagnetismSpinGroup-2026}.

	Although efforts are being made to understand the origin of the chiral magnetic order that characterizes the odd-parity magnet \cite{Yu-Agterberg-OddParityMagnetismDriven-2025,Sim-Rachel-QuantumSpinModels-2026}, much less is known about the mechanism driving the odd-parity spin splitting and its relation to magnetic order.
	In particular, existing models of odd-parity magnets usually treat the electrons as coupled to a static background of fixed local moments, effectively acting as a helical magnetic field \cite{Brekke-Linder-MinimalModelsTransport-2024,Hellenes-Smejkal-PwaveMagnets-2024,Kudasov-Kudasov-TopologicalBandStructure-2024}.
	While useful for exploring the implications of odd-parity spin-splitting, such approaches ignore the rich interplay between electronic correlations and the quantum fluctuations of the local moments.
	Even more concerning is the fact that the minimal models are based on one-dimensional systems \cite{Brekke-Linder-MinimalModelsTransport-2024}, in which long-range magnetic order is precluded by the Hohenberg-Mermin-Wagner theorem \cite{Hohenberg-Hohenberg-ExistenceLongRangeOrder-1967,Mermin-Wagner-AbsenceFerromagnetismAntiferromagnetism-1966}.
	A natural first step in addressing these shortcomings is to promote such models to a strongly correlated setting in which they take the form of one-dimensional Kondo lattices \cite{Doniach-Doniach-KondoLatticeWeak-1977,Tsunetsugu-Ueda-GroundStatePhase-1997,Zachar-Emery-ExactResults1D-1996,Zachar-Tsvelik-OnedimensionalElectronGas-2001}.
     
	In this paper we study a model for a correlated $p$-wave magnet in one dimension by treating it in the framework of Abelian bosonization.
	The motivation for this approach is twofold.
	First, it provides a mapping from a lattice spin model to a quantum field theory in $(1+1)$-dimensional space and time, of which a hallmark is the equivalence between fermions and bosons \cite{Tomonaga-Tomonaga-RemarksBlochsMethod-1950,Luttinger-Luttinger-ExactlySolubleModel-1963,Mattis-Lieb-ExactSolutionManyFermion-1965,Coleman-Coleman-QuantumSineGordonEquation-1975,Mandelstam-Mandelstam-SolitonOperatorsQuantized-1975,Luther-Peschel-CalculationCriticalExponents-1975}.
	This permits treating the fluctuations of the local moments and the itinerant electrons on equal footing.
	Second, the $(1+1)$-dimensional quantum theory can be understood through an equivalent classical statistical field theory in two spatial dimensions.
	This allows us to make precise how true long-range order is traded for quasi-long-range order (QLRO) in one-dimensional condensed matter settings.
    QLRO is the property that correlation functions decay algebraically at large separations. 
	Altogether, this provides a tractable approach for addressing how the $p$-wave spin polarization of electron bands in the noninteracting models persists in a strongly correlated setting.

    The rest of this paper is structured as follows. 
    In Sec.~\ref{sec:field_theory} we derive the bosonized low-energy field theory for the one-dimensional $p$-wave magnet, starting from a lattice model inspired by Ref.~\cite{Brekke-Linder-MinimalModelsTransport-2024}.
    In Sec.~\ref{sec:spin_chain_ground_state} we examine the ground-state correlations of the spin chain to elucidate the connection between the classical ground state assumed in Ref.~\cite{Brekke-Linder-MinimalModelsTransport-2024} and the QLRO that develops in the spin chain.
    In Sec.~\ref{sec:phase_diagram_general} we discuss the effects of the perturbative interaction terms generated in the theory and construct the weak-coupling phase diagram.
    We pay special attention to the appearance of new interaction terms allowed by a commensurability condition on the electronic filling.
    In Sec.~\ref{sec:spectral_function}, we compute the electron spectral function in this special case and demonstrate its pronounced $p$-wave character, despite the absence of electron bands. 
    Finally, we summarize the results and their implications in Sec.~\ref{sec:conclusion}.
    We use units in which $\hbar = k_{B} = 1$.

    \section{From lattice spin Hamiltonian to bosonic field theory}\label{sec:field_theory}

    Here we introduce a microscopic spin lattice model whose classical ground state is the $p$-wave magnet state in Ref.~\cite{Brekke-Linder-MinimalModelsTransport-2024}.  
    We analyze this spin model by bosonizing its Jordan-Wigner representation, identifying the most relevant interaction terms and the low-energy excitations.
    The spin system is coupled to a proximate metal chain and the full theory is bosonized, giving rise to the complete low-energy effective field theory of the $p$-wave magnet.
    Similar approaches have been employed previously to study the Kondo-Heisenberg chain \cite{White-Affleck-DimerizationIncommensurateSpiral-1996,Zachar-Emery-ExactResults1D-1996,Sikkema-White-SpinGapDoped-1997,Zachar-Tsvelik-OnedimensionalElectronGas-2001,Berg-Kivelson-PairDensityWaveCorrelationsKondoHeisenberg-2010}.

    \subsection{Odd-parity Kondo lattice}\label{sec:latticemodel}

\begin{figure}[htb]
        \centering
        \includegraphics[width=\columnwidth]{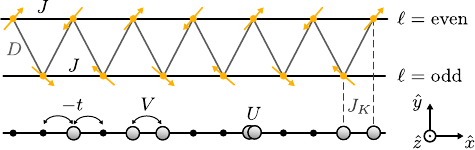}
        \caption{Schematic illustration of the system governed by the Hamiltonian in Eq.~\eqref{eq:H_tot}. 
        The yellow arrows represent the local moments, and the shaded circles represent the itinerant electrons.
        The spin chain is represented as a zigzag ladder, with intrachain nearest-neighbor exchange coupling $J$ and interchain nearest-neighbor Dzyaloshinskii-Moriya coupling $D$ along the $\hat{\v{z}}$ direction.
        The spin chain is coupled to the metal chain through the Kondo coupling $J_{K}$, and the electrons have nearest-neighbor hopping parameter $t$, on-site Hubbard interaction $U$ and nearest-neighbor interaction $V$. 
        }
        \label{fig:model_schematic}
    \end{figure}

    The $p$-wave magnet is described by the following Hamiltonian
    \begin{align}
        H &= H_{\mathrm{OPM}} + H_{\mathrm{EH}} + H_{\mathrm{K}}, \label{eq:H_tot}
    \intertext{where}
        H_{\mathrm{OPM}} &\coloneqq -D \sum_{i}  \hat{\v{z}} \cdot \left( \v{S}_{i} \times \v{S}_{i+1} \right) + J \sum_{i} \v{S}_{i} \cdot \v{S}_{i+2} \label{eq:H_OPM} \\
        \begin{split} 
        H_{\mathrm{EH}} &\coloneqq - t \sum_{\langle i, j \rangle } \sum_{\lambda} c^{\dagger}_{i\lambda} c_{j\lambda}^{\mathstrut} \\
        &\hspace{1.5em}+ U \sum_{i} n_{i\uparrow} n_{i\downarrow}  + V \sum_{\langle i,j\rangle} n_{i} n_{j},
        \label{eq:H_EH}
        \end{split}\\
        H_{\mathrm{K}} &\coloneqq J_{K} \sum_{i} \v{S}_{i} \cdot \v{s}_{i}, \label{eq:H_K}
    \end{align}
    where $\v{S}_{i}$ is a localized spin-$1/2$ degree of freedom obeying the spin algebra $[S^{a}_{i},S^{b}_{j}] = \iu \delta_{ij}\epsilon^{abc} S^{c}_{i}$, and $c_{i\lambda}^{\dagger}$ and $c_{i\lambda}^{\mathstrut}$ are the anticommuting creation and annihilation operators for a fermionic degree of freedom with spin $\lambda=\uparrow,\downarrow$. 
    The particle densities are denoted by $n_{i\lambda} \coloneqq c^{\dagger}_{i\lambda} c^{\mathstrut}_{i\lambda}$ and $n_{i} \coloneqq \sum_{\lambda} n_{i\lambda}$, and the spin density is denoted by $\v{s}_{i} \coloneqq c^{\dagger}_{i\alpha} \bm{\sigma}_{\alpha\beta} c^{\mathstrut}_{i\beta}/2$. 
    The spin model includes nearest-neighbor Dzyaloshinskii-Moriya interaction (DMI) $D$ along the $\hat{\v{z}}$ direction and next-nearest-neighbor antiferromagnetic exchange $J > 0$.
    This  choice is motivated by the fact that the classical ground state of such a model has the desired composite $\mathcal{T}\tau$ symmetry characteristic of odd-parity magnets \cite{Brekke-Linder-MinimalModelsTransport-2024,Hellenes-Smejkal-PwaveMagnets-2024}. 
    A more reasonable model includes the nearest-neighbor exchange $J_1\sum_{\langle i,j\rangle} \v{S}_{i}\cdot \v{S}_{j}$, introducing frustration.
    However, being primarily concerned with how the odd-parity spin splitting derives from the magnetic state, and not how such a state may arise, we defer the inclusion of such terms to future work.
    The metal chain is described by the extended Hubbard (EH) model shown in Eq.~\eqref{eq:H_EH} and includes nearest-neighbor hopping $t$, on-site interaction $U$ and nearest-neighbor interaction $V$. 
    The coupling between the metal chain and the spin chain is described by the Kondo exchange interaction in Eq.~\eqref{eq:H_K} where $J_{K}$ denotes the Kondo coupling.
    The model is dubbed an odd-parity Kondo lattice and is illustrated in Fig.~\ref{fig:model_schematic}.

    We will be concerned with the low-energy and long-wavelength behavior of the theory, obtained by taking the continuum limit of each chain depicted in Fig.~\ref{fig:model_schematic} separately, and linearizing the resulting fermionic dispersions around the Fermi points.
    Such an approximation necessitates restricting to the \textit{weak-coupling limit} defined by $J_K \ll J, t$ and $D \ll J$ \cite{Zachar-Tsvelik-OnedimensionalElectronGas-2001,Zachar-Emery-ExactResults1D-1996}.
    This limit renders the Ruderman-Kittel-Kasuya-Yosida (RKKY) interactions negligible compared to the exchange interactions in Eq.~\eqref{eq:H_OPM} as $J_{\mathrm{RKKY}} \sim J_K^2/t \ll J$ \cite{Tsunetsugu-Ueda-GroundStatePhase-1997}.

    \subsection{Bosonization of the magnetic insulator}\label{sec:bosonization_insulator}

    To proceed with the bosonization of the theory defined by Eq.~\eqref{eq:H_OPM} we view the interacting spin chain as a two-legged zigzag ladder, as illustrated in Fig.~\ref{fig:model_schematic}.
    Let $S^{a}_{j,\ell}$ denote the $a$ component of the spin operator at site $j$ of chain $\ell = 1, 2$, corresponding to odd and even sites respectively. 
    With this notation, the spin Hamiltonian takes the form $H_{\mathrm{OPM}} = H_{J} + H_{D}$, where
    \begin{equation}
    \begin{split}
        H_{\mathrm{OPM}} = J &\sum_{j \, \text{odd}} \v{S}_{j,1} \cdot \v{S}_{j+2,1} + J \sum_{j \, \text{even}} \v{S}_{j,2} \cdot \v{S}_{j+2,2}  \\
        -& D \sum_{j \, \text{odd}} \hat{\v{z}} \cdot \left( \v{S}_{j,1} \times \v{S}_{j+1,2} \right) \\
        -& D \sum_{j \, \text{even}} \hat{\v{z}} \cdot \left( \v{S}_{j,2} \times \v{S}_{j+1,1} \right).
    \end{split}
    \end{equation}
    The spin system is equivalently described in terms of spinless second-quantized fermionic operators $ d_{j,\ell}^{\mathstrut}, d^{\dagger}_{j,\ell} $ with the help of the Jordan-Wigner transformation
    \begin{subequations}\label{eq:JW}
    \begin{equation}
        S^{+}_{j,\ell} = d^{\dagger}_{j,\ell} W_{j,\ell} \quad \text{and} \quad S^{z}_{j,\ell} = d_{j,\ell}^{\dagger} d_{j,\ell}^{\mathstrut} - \frac{1}{2},
    \end{equation}
    where $S^{+}_{j,\ell} = (S^{-}_{j,\ell})^{\dagger}$ and 
    \begin{equation}\label{eq:JW_string}
        W_{j,\ell} \coloneqq \prod_{i < j} \exp( - \iu \pi d_{i,\ell}^{\dagger} d_{i,\ell}^{\mathstrut} ),
    \end{equation}
    \end{subequations}
    denotes the Jordan-Wigner string \cite{Jordan-Wigner-UeberPaulischeAequivalenzverbot-1928}.
    The commutation relations for the spin degrees of freedom imply that the creation and annihilation operators $d^{\dagger}_{j,\ell}$ and $d_{j,\ell}$ obey the canonical anticommutation relations
    \begin{equation}
        \big\{ d_{i,\ell}^{\mathstrut}, d_{j,\ell'}^{\dagger} \big\} = \delta_{\ell\ell'} \delta_{ij}, \quad \text{and}\quad \big\{d_{i,\ell}^{\mathstrut}, d_{j,\ell'}^{\mathstrut} \big\} = 0.
    \end{equation}

    After shifting the momenta of the fermions on each chain by $\pi$ via a gauge transformation, the Fourier-transformed kinetic part of the intrachain Hamiltonian $H_J$ takes the form $H_{J}^{(0)} = \sum_{k,\ell} \varepsilon_{\ell}(k) d^{\dagger}_{\ell}(k) d_{\ell}^{\mathstrut}(k)$, where $\varepsilon_{\ell}(k) = - J \cos(2ka)$ \cite{Giamarchi-Giamarchi-QuantumPhysicsOne-2004}.
    The low-energy sector of this Hamiltonian is described by left- and right-moving fermions of each chain with velocities $v_F^{S} = 2 J a$, where the superscript $S$ will be used to emphasize that a quantity relates to the spin chain.    
    In the continuum limit, we replace the lattice creation and annihilation operators by left- and right-moving fermion fields $\psi_{r,\ell}(x)$ according to 
    \begin{equation}\label{eq:JW_fermions_left_right}
        d_{j,\ell} \to \sqrt{2a} \left[ \e^{\iu k_F^S x_j} \psi_{R, \ell}(x_j) + \e^{-\iu k_F^S x_j} \psi_{L,\ell}(x_j) \right],
    \end{equation}
    where $x_j = a j $, and $j$ is an odd or even integer for $\ell = 1,2$ respectively.
    The bosonized field operators are defined by
    \begin{equation}\label{eq:psi_spins}
        \psi_{r,\ell}(x) = U_{r,\ell} \frac{1}{\sqrt{2\pi \alpha}}\e^{-\iu \left( r \phi_{\ell}(x) - \theta_{\ell}(x) \right) },
    \end{equation}
    where $U_{r,\ell}$ is a Klein-factor and $\alpha$ is a soft momentum cutoff \cite{Giamarchi-Giamarchi-QuantumPhysicsOne-2004}.
    Using this continuum limit, the remaining terms of the exchange Hamiltonian $H_{J}$ combine into two copies of the well-known bosonic field theory of the Heisenberg antiferromagnet \cite{Giamarchi-Giamarchi-QuantumPhysicsOne-2004,Cabra-Pujol-FieldtheoreticalMethodsQuantum-2004} 
    \begin{subequations}\label{eq:H_J_bosonized}
        \begin{align}
            H_{J} =& \frac{1}{2\pi} \int \D x \sum_{\ell}  \left[uK \left( \partial_x \theta_{\ell} \right)^2 + \frac{u}{K} \left( \partial_x \phi_{\ell} \right)^2 \right] \label{eq:H_J_0} \\
            &+ \frac{2 g_1}{(2\pi \alpha)^2} \int \D x \sum_{\ell} \cos( 4 \phi_{\ell}),
    \end{align}
    \end{subequations}
    where $u K = v_F^S$, $ u/K  = v_F^S [ 1 + 4 J a( 1 + a^2 / \alpha^2 )/(\pi v_F^S) ]$ and $ g_1 = 2 J a $.
    Note that a shift $\phi_{2} \mapsto \phi_{2} + \pi/4$ has been performed in order to arrive at this result.

    The DMI term $H_D$ acts as an interchain interaction that couples the two bosonic fields governed by Eq.~\eqref{eq:H_J_bosonized}.
    Using the Jordan-Wigner transformation, one obtains
    \begin{equation}\label{eq:H_D_fermionized}
        \begin{split}
             H_{D} = &\frac{D}{2\iu} \sum_{j \, \text{odd}} \left[ d_{j,1}^{\dagger} W_{j,1}^{\mathstrut} W_{j+1,2}^{\mathstrut} d_{j+1,2}^{\mathstrut} - \hc  \right]  \\
            &- \frac{D}{2\iu} \sum_{j \, \text{even}} \left[ d_{j,2}^{\dagger} W_{j,2}^{\mathstrut} W_{j+1,1}^{\mathstrut} d_{j+1,1}^{\mathstrut} - \hc  \right].
        \end{split}
    \end{equation}
    After some straightforward but tedious algebra (see appendix \ref{app:bosonization_details}) the continuum limit of the DMI term gives rise to the interaction term 
    \begin{equation}\label{eq:H_D_bosonized}
        H_D = \frac{2 g_2}{(2\pi \alpha)^2} \int \D x \, \sin( \theta_2 - \theta_{1}),
    \end{equation}
    where $g_2 = 2\pi \alpha D$.
    Equations \eqref{eq:H_J_bosonized} and \eqref{eq:H_D_bosonized} define the bosonic field theory that describes the spin-sector of the $p$-wave magnet.
 
    \subsection{Bosonization of the odd-parity Kondo lattice}\label{sec:bosonization_full_theory}

    We now turn to the bosonization of the full theory for the odd-parity Kondo lattice.
    The kinetic part of the metal chain is described by the $H_{\mathrm{EH}}^{(0)} = \sum_{k, \lambda} \varepsilon(k) c^{\dagger}_{\lambda}(k) c^{\mathstrut}_{\lambda}(k)$, where $\varepsilon(k) = - 2 t \cos(k a)-\mu$ and the corresponding Fermi momentum and velocity are $k_F = \arccos(-\mu/(2t))/a$ and $v_F = 2 t a \sin(k_F a)$ respectively.
    In the continuum limit, we replace the lattice creation and annihilation operators by left- and right-moving fermion fields $\psi_{r, \lambda}(x)$ according to
    \begin{equation}
        c_{j,\lambda} \to \sqrt{a} \left[ \e^{\iu k_F x_j} \psi_{R,\lambda}(x_j) + \e^{-\iu k_F x_j} \psi_{L,\lambda}(x_j) \right],
    \end{equation}
    where $x_j = j a$ and $j\in\mathbb{Z}$.
    The bosonized field operators are now introduced as 
    \begin{equation}\label{eq:psi_electrons}
        \psi_{r,\lambda}(x) = U_{r,\lambda} \frac{1}{\sqrt{2\pi \alpha}} \e^{-\iu \left( r \varphi_{\lambda}(x) - \vartheta_{\lambda}(x) \right) }.
    \end{equation}
    The details of the construction of the effective bosonic field theory for the extended Hubbard model can be found elsewhere and therefore are not presented here (see, e.g., Refs.~\cite{Giamarchi-Giamarchi-QuantumPhysicsOne-2004,Tsuchiizu-Furusaki-GroundstatePhaseDiagram-2004}).
    Using Eq.~\eqref{eq:psi_electrons} and defining the bosonic fields corresponding to collective charge and spin excitations $\varphi_{\rho/\sigma} \coloneqq (\varphi_{\uparrow} \pm \varphi_{\downarrow})/\sqrt{2}$ respectively, one arrives at 
    \begin{align}\label{eq:H_EH_bosonized}
        H_{\mathrm{EH}} =& \frac{1}{2\pi} \int \D x \sum_{\eta = \rho,\sigma} u_{\eta} \left[ K_\eta (\partial_x \vartheta_{\eta})^2 + \frac{1}{K_\eta} (\partial_x \varphi_{\eta})^2 \right] \notag \\
        &+ \frac{2 g_{1,\sigma}}{(2\pi \alpha)^2} \int \D x \, \cos( \sqrt{8} \varphi_{\sigma}  ),
    \end{align}
    where $g_{1,\sigma} = U a + 2 V a \cos(2k_Fa)$, $u_{\rho} K_{\rho} = u_{\sigma} K_{\sigma} = v_F$ and 
    \begin{subequations}\label{eq:u/Ks}
         \begin{align}
             \frac{u_{\rho}}{K_\rho} &= v_F \left\{ 1 + \frac{a}{\pi v_F} \left[ U + 2V \left( 1 - \cos(2 k_F a) \right) \right] \right\} \\
              \frac{u_{\sigma}}{K_\sigma} &= v_F \left\{ 1 - \frac{a}{\pi v_F} \left[ U + 2V \cos(2k_F a) \right] \right\}.
    \end{align}
    \end{subequations}
    In deriving this we have restricted our attention to a metal away from half filling, which permits discarding the $4 k_F$ umklapp terms that derive from the electronic interactions \cite{Giamarchi-Giamarchi-QuantumPhysicsOne-2004}.
    In particular, this leads to a description of the charge sector of the theory given by a Gaussian model.

    The bosonic theories describing the spin chain and the metal chain are coupled through the Kondo term in Eq.~\eqref{eq:H_K}.
    To see which interaction terms can survive the long-wavelength limit it is instructive to decompose both spin fields in their smooth and staggered parts.
    The electron spin density separates into forward- and back-scattering parts as 
    \begin{equation}
        \v{s}_j =  \v{J}_{R}(x_j) + \v{J}_{L}(x_j) + \v{n}(x_j),
    \end{equation}
    where $\v{J}_{R/L}(x) \coloneqq \psi^{\dagger}_{R/L,\lambda}(x) \bm{\sigma}_{\lambda\lambda'} \psi_{R/L,\lambda'}^{\mathstrut}(x)$ and
    \begin{equation}
        \v{n}(x_j) = \e^{-\iu 2 k_F x_j} \v{n}_R(x_j) + \e^{+\iu 2 k_F x_j} \v{n}_{L}(x_j),
    \end{equation}
    and $\v{n}_{R/L}(x) \coloneqq \psi^{\dagger}_{R/L,\lambda}(x) \bm{\sigma}_{\lambda\lambda'} \psi_{L/R,\lambda'}^{\mathstrut}(x)$.
    In the weak-coupling limit, the spin field associated with the local moments can likewise be expanded in its smooth and staggered (ferromagnetic and antiferromagnetic) parts as (cf. Eqs.~\eqref{eq:bosonized_spins_1} and \eqref{eq:bosonized_spins_2})
    \begin{equation}
        \v{S}_{j} = \v{J}_{R}^{S}(x_j) + \v{J}_{L}^{S}(x_j) + (-1)^{j} \v{n}^{S}(x_j).
    \end{equation}
    The Kondo interaction can now be split into forward-scattering terms ($\sim \normord{\v{J} \cdot \v{J}^{S}}$), back-scattering terms ($\sim \normord{\v{n} \cdot \v{n}^{S}}$) and mixed terms ($\sim \normord{\v{n} \cdot \v{J}^{S}}$ or $\sim \normord{\v{J} \cdot \v{n}^{S}}$) \cite{Zachar-Tsvelik-OnedimensionalElectronGas-2001}.
    When the spin chain and the metal chain are relatively incommensurate, the two latter types of scattering terms are rendered irrelevant in the continuum limit, and only the forward-scattering terms need to be considered.

    However, when the electron filling is commensurate with the staggered parts of the localized moment fields the product $\normord{\v{n} \cdot \v{n}^{S}}$ survives the continuum limit.
    As a consequence, this commensurability condition allows additional back-scattering terms which cannot be ignored.
    Since the spin chain will be shown shortly to develop quasi-long range spiral order with a wave vector of magnitude $|Q| = \pi/(2a)$ (see Sec.~\ref{sec:spin_chain_ground_state}), the commensurability condition can be phrased more precisely as $2 k_F = |Q| $.
    A solution to this matching condition is quarter filling of the electrons.
    Although the quasi-long range order with $|Q| = \pi/(2a)$ is the closest resemblance of the composite $\mathcal{T}\tau$ symmetry defining the $p$-wave magnet in Ref.~\cite{Brekke-Linder-MinimalModelsTransport-2024}, the condition for a $p$-wave spin-polarized electron spectrum in this setting is in fact stronger.
    As we shall see shortly, only under the commensurability condition does the electron spectral density retain the specific $p$-wave features of the noninteracting theory in Ref.~\cite{Brekke-Linder-MinimalModelsTransport-2024}.
    As a result, we will argue that this commensurability condition and the interactions allowed by it constitute the microscopic mechanism for $p$-wave magnetism in one dimension.

    At this stage we pause to make a remark about the connection between the field theory of this magnet and that of its cousin: the $d$-wave altermagnet.
    Interestingly, the mechanism allowing for the back-scattering terms in the bosonic field theory bears some resemblance to the way a new term appears in the nonlinear sigma model describing the insulating $d$-wave altermagnet \cite{Lundemo-Sudbo-QuantumCriticalScaling-2025a}.
    In that case, the term $\sim \v{n} \cdot (\partial_{\tau} \v{n} \times \partial_{x}\partial_{y} \v{n})$ appears in the Euclidean Lagrangian because the Néel field $\v{n}$ is commensurate with a spatially staggered part of the next-nearest-neighbor exchange coupling on the checkerboard lattice, which can be thought of as arising from superexchange via a nonmagnetic site on the Lieb lattice \cite{Brekke-Sudbo-TwodimensionalAltermagnetsSuperconductivity-2023,Lundemo-Sudbo-QuantumCriticalScaling-2025a}. 
    This term was proved to give rise to a $d$-wave nondegeneracy of the magnon bands in the $d$-wave altermagnet \cite{Lundemo-Sudbo-QuantumCriticalScaling-2025a}.
    In a similar fashion, the back-scattering terms will be shown to endow the electron spectral function with a $p$-wave structure in the present case.

    After using the Jordan-Wigner transformation in Eq.~\eqref{eq:JW} together with the bosonized form of the field operators in Eqs.~\eqref{eq:psi_spins} and \eqref{eq:psi_electrons} one can deduce the continuum form of the Kondo term. 
    It includes a quadratic term, a forward-scattering term and a back-scattering term, where it is understood that the latter only applies for $2k_F = |Q|$ as explained above.
    In total, $H_K = H_K^{(0)} + H_f + H_b$, where
    \begin{subequations}\label{eq:H_K_bosonized}
    \begin{align}
        H_K^{(0)} = \frac{J_K a}{\sqrt{2} \pi^2} \int \D x \sum_{\ell} \partial_x \phi_{\ell} \partial_x 
        \varphi_{\sigma},
    \end{align}
    and
    \begin{widetext}
    \begin{align}
        \begin{split}
            H_f &= \frac{2 g_f}{(2\pi \alpha)^2} \int \D x \Big[ \cos( \theta_1 - \sqrt{2} \vartheta_{\sigma} )  \sin( 2 \phi_{1} )  \cos( \sqrt{2} \varphi_{\sigma} ) +  \cos( \theta_2 - \sqrt{2} \vartheta_{\sigma} ) \cos( 2 \phi_{2} ) \cos( \sqrt{2} \varphi_{\sigma} ) \Big]
        \end{split}\\
        \begin{split}
            H_b &= \frac{2 g_b}{(2\pi \alpha)^2} \int \D x \Big[ \cos( \theta_2 - \sqrt{2} \vartheta_{\sigma} ) \cos( \sqrt{2} \varphi_{\rho} ) - \cos( \theta_1 - \sqrt{2} \vartheta_{\sigma} ) \sin( \sqrt{2} \varphi_{\rho} )
            \Big],
        \end{split}
    \end{align}
    \end{widetext}
    \end{subequations}
    where $g_b = g_f = J_K (\pi a \alpha)^{1/2}$.
    The complete bosonic low-energy, long-wavelength field theory describing the odd-parity Kondo lattice is described by the continuum Hamiltonian defined by Eqs.~\eqref{eq:H_J_bosonized}, \eqref{eq:H_D_bosonized}, \eqref{eq:H_EH_bosonized} and \eqref{eq:H_K_bosonized}.  

    \section{Ground-state correlations of the spin chain}\label{sec:spin_chain_ground_state}

    Before studying the complete theory we first consider the spin chain on its own.
    The aim of this section will be to determine the relevance of the coupling constants $g_1$ and $g_2$ under the renormalization group (RG) flow.
    Note that allowing the flow of the effective couplings does not contradict the weak-coupling approximation made on the microscopic coupling constants to derive the effective theory.
    Unless explicitly stated otherwise we assume without loss of generality that $g_2 > 0$ and we use the shorthand notation $r \coloneqq (u\tau, x)$.

    To determine the relevance of the $g_1$ and $g_2$ terms we regard them as perturbations to the free theory defined by $g_1 = g_2 = 0$, shown in Eq.~\eqref{eq:H_J_0}. 
    The corresponding operators $\mathcal{O}_1(r) = \cos(4 \phi_\ell(r))$ and $\mathcal{O}_2(r) = \sin( \theta_2(r) - \theta_1(r) )$ may be expressed in terms of the vertex operators $\e^{\iu n \phi_\ell}$, $\e^{\iu m \theta_\ell}$, with real coefficients $n,m$. 
    Since the theory is Gaussian, the cumulant expansion truncates at second order, so that $\langle \e^{A} \rangle = \e^{\langle A^2 \rangle/2}$.
    This implies that the scaling dimensions of $\mathcal{O}_{1/2}(r)$ can be directly obtained from the Luttinger liquid (LL) correlation functions \cite{Giamarchi-Giamarchi-QuantumPhysicsOne-2004}
    \begin{subequations}\label{eq:LL_correlations}
        \begin{align}
            \left\langle \left[ \phi_{\ell}(r) - \phi_{\ell}(0) \right]^2 \right\rangle &\sim K \log(\frac{\abs{r}}{\alpha}) \\
            \left\langle \left[ \theta_{\ell}(r) - \theta_{\ell}(0) \right]^2 \right\rangle &\sim \frac{1}{K} \log(\frac{\abs{r}}{\alpha}).
        \end{align}
    \end{subequations}
    This yields the equal-time correlation function
    \begin{equation}
        \langle \mathcal{O}_{1/2}(x) \mathcal{O}_{1/2}(0) \rangle \sim \frac{1}{\abs{x}^{2 \Delta_{1,2}}},
    \end{equation}
    where $\Delta_1 = 4 K$ and $\Delta_2 = 1/(2K)$. 
    At the LL fixed point $K=1/2$, $\Delta_1 = 2$ and $\Delta_2 = 1$. 
    Hence, the $\mathcal{O}_1$ perturbation is marginal while the $\mathcal{O}_2$ perturbation is relevant.
    Based on this scaling analysis, we let $g_1=0$ and assume that the flow of $g_2$ will dominate the long-wavelength physics.
    As $g_2$ flows to strong coupling it is clear that the antisymmetric theta field $\theta_{a} \coloneqq (\theta_1 - \theta_2)/\sqrt{2}$ will be pinned to the minima of $\sin( - \sqrt{2} \theta_a )$, while the symmetric sector defined by $\theta_{s} \coloneqq (\theta_1 + \theta_2)/\sqrt{2}$ retains its description as a LL.
    Expanding in small fluctuations $\tilde{\theta}_a$ about the minimum configurations, the Euclidean action of the antisymmetric sector reads
    \begin{equation}
        S_a = \frac{K}{2\pi} \int \D \tau \D x \left[ \frac{1}{u} (\partial_{\tau} \tilde{\theta}_a )^2 + u ( \partial_x \tilde{\theta}_a )^2 + m^2_a \tilde{\theta}_a^2  \right],
    \end{equation}
    where the mass of these fluctuations is given by $m_a^2 = 4\pi g_2 /((2\pi \alpha)^2 K)$.
    In the following we will omit the tilde for brevity.
    We note that the precise character of the strong-coupling fixed point cannot be deduced from the present perturbative analysis. 
    However, the following analysis requires only that $\theta_a$ acquires a mass.
    
    To make further connections between this theory and the classical picture in Fig.~\ref{fig:model_schematic} and in Ref.~\cite{Brekke-Linder-MinimalModelsTransport-2024} we consider the spin correlations when $\theta_a$ becomes massive. 
    Since the local spin fields with $\ell=1,2$ reside on odd and even sites respectively, their bosonized forms are not simply related by a relabelling $1 \leftrightarrow 2$. 
    In particular, we find
    \begin{subequations}\label{eq:bosonized_spins_1}
        \begin{align}
            S_{1}^{z}(x) &= - \frac{1}{\pi} \partial_x \phi_1(x) + \frac{(-1)^x}{\pi \alpha} \sin (2 \phi_{1}(x)) \\
            S_{1}^{+}(x) &= \frac{\e^{-\iu \theta_1(x)}}{\sqrt{2\pi\alpha}} \Big[ (-1)^{x} + \sin( 2 \phi_1(x) ) \Big],
        \end{align}
    \end{subequations}
    and
    \begin{subequations}\label{eq:bosonized_spins_2}
        \begin{align}
            S_{2}^{z}(x) &= - \frac{1}{\pi} \partial_x \phi_2(x) + \frac{(-1)^x}{\pi \alpha} \cos (2 \phi_{2}(x)) \\
            S_{2}^{+}(x) &= \frac{\e^{-\iu \theta_2(x)}}{\sqrt{2\pi\alpha}} \Big[ (-1)^{x} + \cos( 2 \phi_{2}(x) ) \Big].
        \end{align}
    \end{subequations}
    Briefly neglecting the quantum nature of the spin, one may view the fields $2 \phi_{\ell}$ and $\theta_{\ell}$ as the polar and azimuthal angles of the classical spin $\v{S}_{\ell}$ \cite{Giamarchi-Giamarchi-QuantumPhysicsOne-2004}. 
    With this interpretation, the pinning of $\theta_1 - \theta_2 = \pi/2$ corresponds to locking the spins of chain $1$ to be orthogonal to those on chain $2$.
    This coincides with the classical picture displayed in Fig.~\ref{fig:model_schematic}.
    We emphasize that this ordering is not in conflict with the Hohenberg--Mermin--Wagner theorem \cite{Hohenberg-Hohenberg-ExistenceLongRangeOrder-1967,Mermin-Wagner-AbsenceFerromagnetismAntiferromagnetism-1966}; the spin-rotation symmetry is explicitly broken by the presence of the DMI.
    
    To make this classical intuition more precise, we now demonstrate that the spin correlations capture the antiferromagnetic order within each chain and the orthogonal locking between the two chains.
    Using the expressions in Eqs.~\eqref{eq:bosonized_spins_1} and \eqref{eq:bosonized_spins_2} we find that the dominant part of the in-plane intrachain spin correlations is given by the antiferromagnetic part
    \begin{equation}
        \langle S^{+}_{\ell}(x) S_{\ell}^{-}(0) \rangle \sim (-1)^{x} \left( \frac{\alpha}{\abs{x}} \right)^{1/(4K)},
    \end{equation}
    while the ferromagnetic correlations are exponentially suppressed. 
    Likewise, we find that the out-of-plane spin correlations decay \textit{faster} than the in-plane antiferromagnetic correlations for $K=1/2$ 
    \begin{equation}
        \langle S^{z}_{\ell}(x) S_{\ell}^{z}(0) \rangle \sim \left( \frac{\alpha}{\abs{x}} \right)^{2}.
    \end{equation}
    This is the sense in which the system prefers in-plane antiferromagnetic order within each chain although there is no true long-range order effecting it.

    The orthogonal locking of the spins on the odd and even sites can be understood as a nonzero expectation value $\langle \bm{\chi}_{ij} \rangle $ of the vector spin chirality $\bm{\chi}_{ij} \coloneqq \v{S}_i \times \v{S}_j $,
    where $ij$ are neighboring lattice sites.
    By considering the local operator $\chi^{z}(x) \coloneqq (\v{S}_{1}(x) \times \v{S}_2(x))^{z}$ we examine the chiral order characterized by the strong-coupling limit.
    We find that
    \begin{align}
        \left\langle \chi^{z}(x) \right\rangle &\sim \left\langle \sin(\sqrt{2} \theta_a(x) )  \right\rangle \sim \mathrm{const},
    \end{align}
    as $\theta_a$ is massive. 
    This is a result of the explicit symmetry breaking effected by the DMI.
    Together with the dominant antiferromagnetic in-plane correlations, these ground state expectation values exhibit a chiral order resembling the classical ground state, corresponding to quasi-long-range spiral order with wave vector $Q=\pi/(2a)$.
    
    \section{Weak-coupling phase diagram}\label{sec:phase_diagram_general}

    The complete odd-parity Kondo lattice model is now studied in the weak-coupling limit. 
    We will first determine the relevance of the coupling constants $\{ g_{1,\sigma}, g_{1,\ell}, g_2, g_{f},g_b\}$ by regarding the corresponding operators as perturbations to the free theory. 
    The computation of the scaling dimension of $\mathcal{O}_2$ is not altered by introducing the Kondo coupling.
    As a result the operator $\mathcal{O}_2$ has scaling dimension $\Delta_2 = 1$ and is strongly relevant.

    At the Gaussian fixed point the theory is described by a multichannel LL \cite{Kagalovsky-Yurkevich-LocalImpurityMultichannel-2017} with Euclidean Lagrangian 
    \begin{equation}
        \mathcal{L} = \frac{1}{2\pi}  \left( \bm{\phi}^{\mathsf{T}} \,\, \bm{\theta}^{\mathsf{T}} \right) \left[ \tau^{1} \iu \partial_{\tau} - \begin{pmatrix}
            \mathsf{V}_{\phi} & \\
            & \mathsf{V}_{\theta}
        \end{pmatrix} \partial_x \right] \partial_x \begin{pmatrix}
            \bm{\phi} \\
            \bm{\theta}
        \end{pmatrix} 
    \end{equation}
    where $\tau^1 \coloneqq \mathrm{adiag}(\mathds{1}_4,\mathds{1}_4)$ and the vector fields are defined by $\bm{\phi} \coloneqq \left( \varphi_{\rho}, \varphi_{\sigma}, \phi_{s}, \phi_{a} \right)^{\mathsf{T}}$ and $\bm{\theta} \coloneqq \left( \vartheta_{\rho}, \vartheta_{\sigma}, \theta_{s}, \theta_{a} \right)^{\mathsf{T}}$, and the matrices $\mathsf{V}_{\phi}$ and $\mathsf{V}_{\theta}$ are given by 
    \begin{subequations}
        \begin{align}
            \mathsf{V}_{\phi} &= \begin{pmatrix}
                u_{\rho}/K_{\rho} & 0 & 0 & 0 \\
                0 & u_{\sigma}/K_{\sigma} & J_K a /\pi & 0 \\
                0 & J_K a /\pi & u_S/K_S & 0 \\
                0 & 0 & 0 & u_S/K_S
            \end{pmatrix} \\
            \mathsf{V}_{\theta} &= \begin{pmatrix}
                u_{\rho} K_{\rho} & 0 & 0 & 0 \\
                0 & u_{\sigma} K_{\sigma} & 0 & 0 \\
                0 & 0 & u_S K_S & 0 \\
                0 & 0 & 0 & u_S K_S
            \end{pmatrix}.
        \end{align}
    \end{subequations}
    \begin{figure}[htb]
        \centering
        \includegraphics[width=\linewidth]{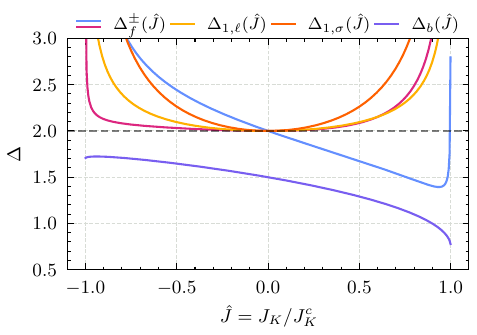}
        \caption{Scaling dimensions of the operators $\mathcal{O}_{f,\pm}$, $\mathcal{O}_{1\ell}$, $\mathcal{O}_{1\sigma}$ and $\mathcal{O}_{b}$ as a function of $\hat{J}\coloneqq J_K/J_{K}^{c}$ for $U=V=0$ and $\mu/t = 0.3$. The dashed gray line indicates the crossover below which the corresponding operator is relevant in the RG sense.
        }
        \label{fig:scaling_dimensions}
    \end{figure}
    
    The scaling dimensions of the remaining operators are determined by casting them in the form of vertex operators $\mathcal{O}(x) = \exp( \iu \v{A}^{\mathsf{T}} \bm{\phi}(x) + \iu \v{B}^{\mathsf{T}} \bm{\theta}(x))$, and subsequently computing the anomalous dimension of the correlation functions $R_{\mathcal{O}}(x) = \langle \mathcal{O}^{\dagger}(x) \mathcal{O}(0) \rangle \sim 1/\abs{x}^{2 \Delta_{\mathcal{O}}}$, where $2 \Delta_{\mathcal{O}} = \eta_{\mathcal{O}}$ and
    \begin{equation}\label{eq:eta_O}
        \eta_{\mathcal{O}} = \frac{1}{2} \v{A}^{\mathsf{T}} \mathsf{K} \v{A} + \frac{1}{2} \v{B}^{\mathsf{T}}\mathsf{K}^{-1} \v{B}.
    \end{equation}
    Here $\mathsf{K} = \mathsf{M} \mathsf{M}^{\mathsf{T}}$ denotes the Luttinger matrix and $\mathsf{M}$ is the matrix that simultaneously diagonalizes $\mathsf{V}_{\phi}$ and $\mathsf{V}_{\theta}$ while preserving the commutators of $\bm{\phi}$ and $\bm{\theta}$ \cite{Kagalovsky-Yurkevich-LocalImpurityMultichannel-2017}.
    Specifically, by introducing the fields $\tilde{\bm{\phi}} = \mathsf{M}^{-1} \bm{\phi}$ and $\tilde{\bm{\theta}} = \mathsf{M}^{\mathsf{T}} \bm{\theta}$ the Hamiltonian is diagonalized as $\mathsf{M}^{\mathsf{T}} \mathsf{V}_{\phi} \mathsf{M} = \mathsf{M}^{-1} \mathsf{V}_{\theta} (\mathsf{M}^{-1})^{\mathsf{T}} = \mathsf{u}$, where $\mathsf{u} = \mathrm{diag}(u_1,\dots,u_4)$ is the diagonal velocity matrix \cite{Kagalovsky-Yurkevich-LocalImpurityMultichannel-2017}. 
    The Luttinger matrix is parametrized by
    \begin{equation}\label{eq:K_total}
        \mathsf{K} = \begin{pmatrix}
            K_{\rho} & 0 & 0 & 0\\
            0 & K_{11} & K_{12} & 0 \\
            0 & K_{21} & K_{22} & 0\\
            0 & 0 & 0 & K_S
        \end{pmatrix}.
    \end{equation}
    Since the antisymmetric sector and the charge sector decouple from the remaining sectors, the problem reduces to an effective two-channel problem.
    This permits using the exact solutions for the two-channel case of Ref.~\cite{Kagalovsky-Yurkevich-LocalImpurityMultichannel-2017} to find the reduced Luttinger matrix $\mathsf{K}'$ corresponding to the center $2\times 2$ block of Eq.~\eqref{eq:K_total}. 
    This yields
    \begin{equation}
        \mathsf{K}'  = \frac{1}{R} \begin{pmatrix}
            K_{\sigma} (u_{\sigma} + u_{S} \gamma) & - \sqrt{K_{\sigma} K_{S}} \tilde{U} \gamma \\
            - \sqrt{K_{\sigma} K_{S}} \tilde{U} \gamma & K_{S} (u_{S} + u_{\sigma} \gamma)
        \end{pmatrix},
    \end{equation}
    where we have introduced $R \coloneqq \sqrt{ u_{\sigma}^2 + u_{S}^2 + 2 u_{\sigma} u_{S}/\gamma }$, $\tilde{U} \coloneqq J_{K} a\sqrt{K_{S} K_{\sigma}}/\pi$ and $\gamma^{-1} \coloneqq \sqrt{1 - \tilde{U}^2/(u_{\sigma} u_{S})}$.
    
    The scaling dimensions of the operators are given by
    \begin{subequations}
        \begin{align}
            &\Delta_{1, \sigma } = 2 K_{11} \\
            &\Delta_{1, \ell} = 2 K_{22} + 2 K_{S} \\
            \begin{split}
                &\Delta_{f}^{\pm} = \frac{1}{4} \bigg[ \frac{2K_{22} + 2K_{12} + K_{11}/2}{K_{11} K_{22} - K_{12}^2} + \frac{1}{2 K_S} \\
                &\qquad\qquad +  2 K_{11} + 2 K_{22} \pm 4 K_{12} + 2 K_{S} \bigg]  
            \end{split}\\
            \begin{split}
                &\Delta_{b}^{\pm} = \frac{1}{4} \bigg[ \frac{2 K_{22} + 2K_{12} + K_{11}/2 }{K_{11} K_{22} - K_{12}^2} +\frac{1}{2 K_S} + 2K_{\rho} \bigg].
            \end{split}
        \end{align}
    \end{subequations}
    The scaling dimensions are plotted in Fig.~\ref{fig:scaling_dimensions} as a function of $J_{K}/J_{K}^c$, where $J_{K}^c \coloneqq \pi \sqrt{u_{\sigma} u_{S} /(K_{\sigma} K_{S}) }/a $.
    Here we have for simplicity set $U = V = 0$ since the picture is not qualitatively altered for nonzero interaction strengths.
    At the ``Toulouse point" $J_K = \pm J_{K}^{c}$ the matrix $\mathsf{K}$ is singular, reflecting the assumption of weak coupling required to derive this theory \cite{Zachar-Tsvelik-OnedimensionalElectronGas-2001}.
    Away from quarter filling it is evident that all interactions except for the forward scattering are irrelevant. 
    If $J_K >0$ forward scattering is relevant, which is consistent with previous studies of the Kondo-Heisenberg lattice \cite{Zhang-Vishwanath-PairdensitywaveSuperconductorDoping-2022}.
    
    The operator that derives from the DMI is nevertheless the most relevant one, suggesting that a more reasonable approach would be to consider all remaining operators as perturbations from the massive theory where $g_2$ runs to strong coupling.
    When $\theta_a$ acquires a mean value in this limit, the dual field $\phi_a$ is disordered.
    This ultimately renders the forward-scattering terms irrelevant, being suppressed by the presence of the field $\phi_a$. 
    The back-scattering terms, however, do not depend on the disordered field.
    Accordingly, the fate of the long-wavelength theory at commensurate filling will be determined by a new strong-coupling fixed point in which $g_b$ also runs to strong coupling.
    We will refer to this as the commensurate phase.
    Conversely, away from commensurate filling, the system is simply described by a multichannel LL.
    In the following, we treat these two cases separately.

    \subsection{Weak-coupling phase diagram away from commensurability}\label{sec:phase_diagram_not_quarter_filling}
    
    \begin{figure*}[htb]
        \centering
        \includegraphics[width=\linewidth]{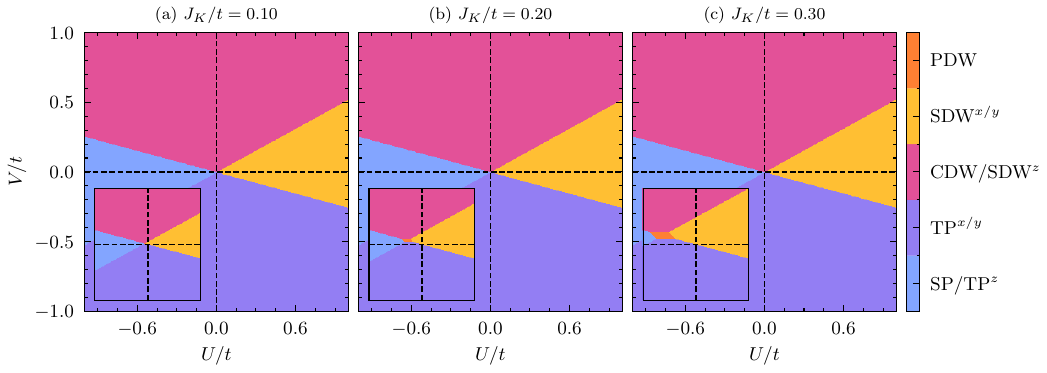}
        \caption{Weak-coupling phase diagram in terms of the on-site interaction $U$ and the nearest-neighbor interaction $V$ for three values of the Kondo coupling $J_K$. 
        The dashed lines indicate the $U = 0$ and $V=0$ axes.
        The phase diagram is computed using $ J/ t = 1$ and $\mu/t = 0.3$.
        The insets show the center of the phase diagram enlarged by a factor $3\times10^{-3}$.
        }
        \label{fig:phase_diagram_Js}
    \end{figure*}    

    Away from commensurate filling the theory is described by the multichannel LL with Euclidean Lagrangian
    \begin{equation}
    \begin{split}
        \mathcal{L} = &\frac{1}{2\pi}  \left( \bm{\phi}^{\mathsf{T}} \,\, \bm{\theta}^{\mathsf{T}} \right) \left[ \tau^{1} \iu \partial_{\tau} - \begin{pmatrix}
            \mathsf{V}_{\phi} & \\
            & \mathsf{V}_{\theta}
        \end{pmatrix} \partial_x \right] \partial_x \begin{pmatrix}
            \bm{\phi} \\
            \bm{\theta}
        \end{pmatrix}  \\
        &+ \frac{1}{2\pi} m_a^2 \theta_a^2,
    \end{split}
    \end{equation}
    where the matrices $\mathsf{V}_{\phi}$ and $\mathsf{V}_{\theta}$ are defined as before.
    The phase diagram of the model is established by computing the scaling dimensions of operators corresponding to charge-density-wave (CDW) order, spin-density-wave (SDW) order, singlet pairing (SP), triplet pairing (TP) and $2k_F$-modulated singlet pairing (PDW) within this theory.
    We characterize the dominating order of the system as the one with the most slowly decaying correlations.
    Since all these operators can be expressed as vertex operators in the form $\mathcal{O}(x) = \exp( \iu \v{A}^{\mathsf{T}} \bm{\phi}(x) + \iu \v{B}^{\mathsf{T}} \bm{\theta}(x))$, the corresponding scaling dimensions are computed straightforwardly by computing the anomalous dimension of the correlation functions $R_{\mathcal{O}}(x) = \langle \mathcal{O}^{\dagger}(x) \mathcal{O}(0) \rangle \sim 1/\abs{x}^{\eta_{\mathcal{O}}}$ as in Eq.~\eqref{eq:eta_O}.
    Since the antisymmetric $\theta$ field is gapped and the charge sector does not couple to the spin and symmetric sectors, the problem reduces again to an effective two-channel problem in the spin and symmetric sectors.

    The operators we consider are given by \cite{Giamarchi-Giamarchi-QuantumPhysicsOne-2004}
    \begin{subequations}
        \begin{align}
            &\mathcal{O}_{\mathrm{SP}/\mathrm{TP}^{z}}(x) \sim \psi^{\dagger}_{R,\uparrow} \psi^{\dagger}_{L,\downarrow} \pm \psi^{\dagger}_{L,\uparrow} \psi^{\dagger}_{R,\downarrow} \label{eq:O_SP} \\
            &\mathcal{O}_{\mathrm{TP}^{x/y}}(x) \sim \psi^{\dagger}_{R,\uparrow} \psi^{\dagger}_{L,\uparrow} \pm \psi^{\dagger}_{L,\downarrow} \psi^{\dagger}_{R,\downarrow} \label{eq:O_TP_xy} \\
            &\mathcal{O}_{\mathrm{CDW}/\mathrm{SDW}^{z}}(x) \sim \psi^{\dagger}_{R,\uparrow} \psi^{\mathstrut}_{L,\uparrow} \pm \psi^{\dagger}_{R,\downarrow} \psi^{\mathstrut}_{L,\downarrow} \label{eq:O_CDW} \\
            &\mathcal{O}_{\mathrm{SDW}^{x/y}}(x) \sim \psi^{\dagger}_{R,\uparrow} \psi^{\mathstrut}_{L,\downarrow} \pm \psi^{\dagger}_{R,\downarrow} \psi^{\mathstrut}_{L,\uparrow} \label{eq:O_SDW_xy}\\
            &\mathcal{O}_{\mathrm{PDW}}(x) \sim \e^{-2\iu k_F x} \psi^{\dagger}_{R,\uparrow} \psi^{\dagger}_{R,\downarrow} + \e^{2\iu k_F x}  \psi^{\dagger}_{L,\uparrow} \psi^{\dagger}_{L,\downarrow}, \label{eq:O_PDW}
        \end{align}
    \end{subequations}
    where it is understood that the fermion fields on the right-hand side of these equations are evaluated at $x$.
    Note that one can write down triplet PDW operators as well.
    However, $z$-triplet pairing has the same decay exponent as the singlet pairing and the $x/y$ triplets are subdominant with respect to all the other orders we consider.
    We therefore restrict attention to only the PDW operator in Eq.~\eqref{eq:O_PDW}.
    These operators have decay exponents that can be expressed in terms of components of the Luttinger matrix as follows 
    \begin{subequations}
        \begin{align}
            \eta_{\mathrm{SP}/\mathrm{TP}^{z}} &= K_{11} + \frac{1}{K_{\rho}} \\
            \eta_{\mathrm{TP}^{x/y}} &= \frac{K_{22}}{K_{11} K_{22} - K_{12}^2} + \frac{1}{K_{\rho}} \\
            \eta_{\mathrm{CDW}/\mathrm{SDW}^{z}} &= K_{11} + K_{\rho} \phantom{\frac{1}{2}} \\
            \eta_{\mathrm{SDW}^{x/y}} &= \frac{K_{22}}{K_{11} K_{22} - K_{12}^2} + K_{\rho} \\
            \eta_{\mathrm{PDW}} &= K_{\rho} + \frac{1}{K_{\rho}}.
        \end{align}
    \end{subequations}
    
    The resulting phase diagram is shown in Fig.~\ref{fig:phase_diagram_Js}.
    The phase diagram is most easily understood as a perturbation of the standard EH model \cite{Sengupta-Campbell-BondorderwavePhaseQuantum-2002,Tsuchiizu-Furusaki-GroundstatePhaseDiagram-2004,Penc-Mila-PhaseDiagramOnedimensional-1994}.
    The oblique lines that separate the four most prominent phases in Fig.~\ref{fig:phase_diagram_Js} are defined by the point at which interactions become effectively repulsive or attractive in the spin and charge channels.
    For attractive interactions $U,V < 0$, the dominant correlations are generally superconducting and for repulsive interactions $U,V >0$, the dominant correlations are of the density-wave type.
    This phase boundary is defined by the point at which $K_{\rho} = 1$, i.e.,
    \begin{equation}
        U + 2 V ( 1 - \cos( 2 k_F a ) ) = 0.
    \end{equation}
    Moreover, there is a crossover from SP to TP and CDW to SDW orders when the interactions become repulsive in the spin channel. 
    This phase boundary is found at $K_{11} = K_{22} /(K_{11} K_{22} - K_{12}^2 )$, which can be expanded to leading order in $J_K$ to give
    \begin{equation}
        U + 2 V \cos( 2 k_F a  ) = -\frac{K_S a}{(v_F + u_S) \pi} J_K^2 + \mathcal{O}(J_K^4).
    \end{equation}
    At half filling and $J_K = 0$ we obtain the well-known $U \approx 2 V$ crossover between CDW and SDW order in the one-dimensional EH model at half filling \cite{Sengupta-Campbell-BondorderwavePhaseQuantum-2002,Tsuchiizu-Furusaki-GroundstatePhaseDiagram-2004}.
    Notably, the Kondo coupling $J_K$ displaces this phase boundary along the one defined by $K_{\rho} = 1$.
    At the intersection between the density-wave and superconducting regions of the phase diagram we now find an extended parameter regime where the PDW correlations dominate.
    As $J_K$ is increased the extent of this region increases, suggesting that the Kondo coupling can stabilize the PDW phase.
    This scenario is insensitive to the sign of $J_K$.

    In the standard Kondo-Heisenberg lattice, the PDW state has been found to dominate for antiferromagnetic Kondo coupling \cite{Berg-Kivelson-PairDensityWaveCorrelationsKondoHeisenberg-2010, Zachar-Tsvelik-OnedimensionalElectronGas-2001}.
    This is to be contrasted with the present case, where it appears in a narrow parameter regime in the weak-coupling limit.
    This difference derives directly from the strongly relevant operator $\mathcal{O}_2$ that renders the forward-scattering terms irrelevant at strong coupling $g_2\to\infty$.

    The presence of a gapped field in the theory can promote composite orders to compete with purely electronic correlations.
    In our case, the gapped $\theta_a$ field reduces the decay exponent of certain order parameters involving the spin-polaron, defined by
    \begin{equation}
        \psi_{\mathrm{p}\lambda}(x) = \frac{1}{2} \sum_{\lambda'} \left( \v{S}(x) \cdot \bm{\sigma}_{\lambda\lambda'} \right) \psi_{\lambda'}(x).
    \end{equation}
    In particular, we find that singlet and $z$-triplet pairing ($\mathrm{SP}/\mathrm{TP}^{z}$) of electrons and spin-polarons have the same decay exponent, whereas the decay exponent of polaron $x/y$-triplet pairings is larger than the corresponding one for electrons.
    The $2k_F$-modulated pairings of spin-polarons have the same decay exponent as the corresponding ones of electrons.
    Moreover, charge-density wave and $z$-spin-density wave ordering of electrons and spin-polarons have the same decay exponent, whereas the $x/y$ spin-density wave correlations of spin-polarons are subleading compared to those of the electrons.  
    To determine which of the orders dominate when the decay exponents coincide one has to compute logarithmic corrections to scaling or compare the amplitudes of the correlation functions \cite{Zhang-Vishwanath-PairdensitywaveSuperconductorDoping-2022}.
    We defer such an analysis to future work.

    \subsection{Weak-coupling phase diagram at commensurability}\label{sec:phase_diagram_quarter_filling}
    
    We now consider the specific case of commensurate filling.
    This modifies the scaling analysis that laid the basis for the weak-coupling phase diagram of the previous section.
    In particular, we now consider the commensurate phase and proceed in an analogous manner to the incommensurate case, identifying first the pinned and disordered fields.

    In the commensurate phase the antisymmetric $\theta$ field is pinned to the minimum of $g_2 \sin( - \sqrt{2} \theta_{a})$, which depends on the sign of $g_2 \propto D$.
    That is, $\sqrt{2} \theta_{a} = \chi \pi/2 $ where $\chi = \sign{D}$, thus establishing quasi long-range spiral order with wave vector $Q=\chi \pi/(2a)$ as discussed in Sec. \ref{sec:spin_chain_ground_state}.
    Inserting this into the back-scattering terms yields
    \begin{equation}
        \mathcal{O}_b = \chi \cos( 3 \Phi^{\chi}_{1}/\sqrt{2} ) + \sin( 3 \Phi^{\chi}_1/\sqrt{2} ),
    \end{equation}
    where $\Phi^{\chi}_{1} = ( \theta_s - 2 \vartheta_{\sigma} - 2\chi \varphi_{\rho} )/3$.
    After pinning $\theta_a$ and $\Phi_{1}^{\chi}$ to their respective minima, the remaining critical modes of the theory are described by the orthogonal fields $\Phi_{2}^{\chi} = ( \vartheta_{\sigma} - \chi \varphi_{\rho} )/\sqrt{2}$ and $\Phi^{\chi}_{3} = ( \vartheta_{\sigma} + 4 \theta_s + \chi \varphi_{\rho} )/(3\sqrt{2})$.
    As a result, the problem again reduces to an effective two-channel problem and one can find the new Luttinger matrix $\mathsf{K}$ analytically.

    In the commensurate phase, the composite field $\Phi_{1}^{\chi}$ is pinned in addition to the antisymmetric field $\theta_{a}$, meaning that the dominant order is characterized by a finite expectation value of
    \begin{equation}\label{eq:pinned_phi_1}
        \cos( \frac{3}{\sqrt{2}} \Phi_{1}^{\chi} ) \sim S^{+} \psi_{-\chi,\uparrow}^{\dagger} \psi_{\chi,\downarrow}^{\mathstrut} + \hc,
    \end{equation}
    where $\psi_{\chi = +,\lambda} \coloneqq \psi_{R,\lambda}$ and $\psi_{\chi = -,\lambda} \coloneqq \psi_{L,\lambda}$.
    The order that develops upon pinning $\Phi_{1}^{\chi}$ is a composite SDW distinct from the SDW orders considered earlier.
    We interpret this as the electrons forming a SDW commensurate with the chiral background of the spin chain, forming spin singlets with the local moments, through an effect analogous to the Kondo effect \cite{Doniach-Doniach-KondoLatticeWeak-1977}.
    Here, however, the effect is insensitive to the sign of the Kondo coupling and the chirality of the spin chain $\chi$ selects electrons only with a specific helicity $\eta \coloneqq \lambda r$ to form singlets with the local moments.
    Specifically, only the pair of electrons with helicity opposite of the chirality of the spin chain $\eta = -\chi$ form singlets, while the other pair is free to propagate.
    This mechanism is analogous to the asymmetrical scattering mechanism responsible for nonreciprocal transport in chiral magnets \cite{Ishizuka-Nagaosa-AnomalousElectricalMagnetochiral-2020, Murata-Ishizuka-KondoEffectNonreciprocal-2026}.
    The commensurate phase thus realizes a helical Luttinger liquid.
    In the following we will elucidate what imprint it leaves on the single-particle spectrum. 

    \section{Electron spectral function}\label{sec:spectral_function}

    \begin{figure*}[htb]
    \centering
    \includegraphics[width=\textwidth]{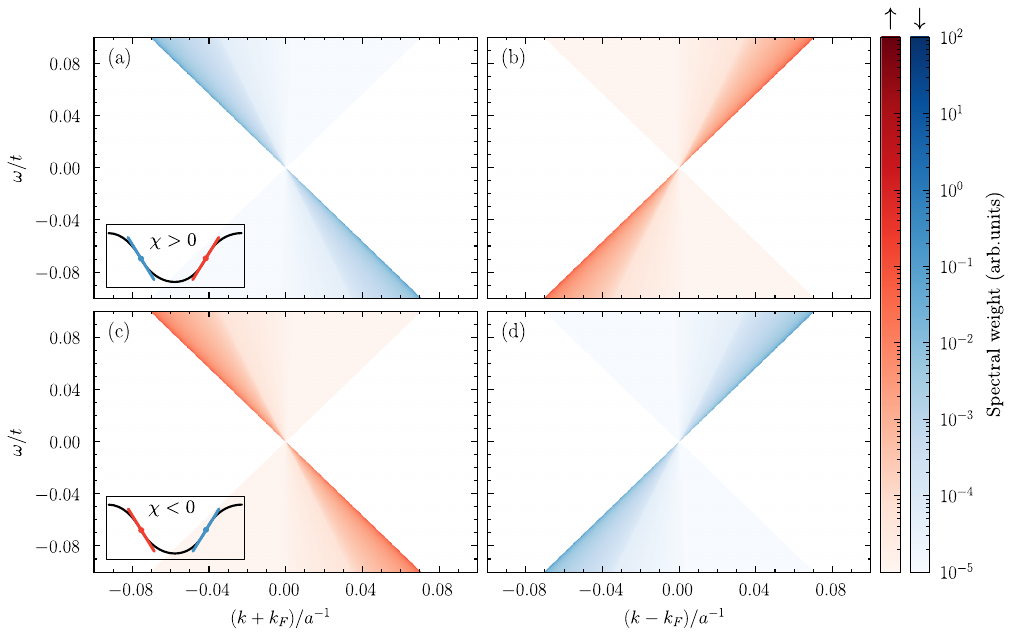}
    \caption{Electron spectral function at quarter filling for (a-b) positive chirality of the spin chain $\sqrt{2}\theta_a=\pi/2$, and (c-d) negative chirality $\sqrt{2}\theta_a=-\pi/2$.
    Red and blue indicate the contribution to the total spectral weight from the spin up and down electrons, respectively. 
    Computed for $U=V=0$, $J_K/t=0.3$.
    Panels (a) and (c) show the spectral function for momenta close to the left Fermi point $-k_F$, and panels (b) and (d) show the spectral function for momenta close to the right Fermi point $k_F$.
    This is illustrated by the insets showing the corresponding points on top of the free-electron dispersion.
    }
    \label{fig:spectral_function}
\end{figure*}

    Finally we consider the electron Green's function and the corresponding spectral density away from and at commensurate filling. 
    In the noninteracting models for $p$-wave magnets the defining feature is the electron band with $p$-wave spin polarization \cite{Brekke-Linder-MinimalModelsTransport-2024,Hellenes-Smejkal-PwaveMagnets-2024}.
    In strongly interacting systems the notion of electron bands is obfuscated. 
    The effect is particularly strong in one dimension, where the spin and charge degrees of freedom of the electron separate \cite{Giamarchi-Giamarchi-QuantumPhysicsOne-2004}.
    However, one can still look for remnants of the $p$-wave polarized bands in the electron spectral function.
    This is the aim of what follows. 

    The single-electron spectral function is defined by
    \begin{equation}
        A_{r\lambda}(q,\omega) = - \frac{1}{\pi} \Im \int \D x \, \D t \, \e^{-\iu (k x - \omega t)} G_{r\lambda}^{\mathrm{ret}}(x,t),
    \end{equation}
    where $k = r k_F + q$ 
    and 
    \begin{equation}
        G_{r\lambda}^{\mathrm{ret}}(x,t) \coloneqq - \iu \theta(t) \big\langle \{ \psi_{r,\lambda}^{\mathstrut}(x,t), \psi_{r,\lambda}^{\dagger}(0,0) \} \big\rangle,
    \end{equation}
    is the retarded electron Green's function and $\theta(t)$ denotes the Heaviside step function.
    By writing
    \begin{equation}
        G^{\mathrm{ret}}_{r\lambda}(x,t) = - \iu \theta(t) \left[ G_{r\lambda}(x,t) + G_{r\lambda}(-x,-t) \right],
    \end{equation}
    we can obtain the retarded Green's function by computing $G_{r\lambda}(x,\tau) \coloneqq - \langle T_{\tau} \psi_{r,\lambda}^{\mathstrut}(x,\tau) \psi_{r,\lambda}^{\dagger}(0,0) \rangle $ and analytically continuing to real time $\tau \to \iu t + \sign(t) 0^{+}$.
    Away from commensurate filling, the electron Green's functions are spin-degenerate and inversion symmetric. 
    The spectral function inherits these symmetries and will consequently not show any remnants of the odd-parity spin polarization away from commensurate filling.
    
    In the commensurate phase the result is very different.
    For positive chirality of the spin chain $\chi = +$ the electron operators are 
    \begin{subequations}
            \begin{align}
                &\psi_{+ \uparrow}(x,\tau) \sim \e^{-\iu \sqrt{2} ( \Theta^{+}_{2} - \Phi_{2}^{+} )} \\
                &\psi_{+ \downarrow}(x,\tau) \sim \e^{-\iu ( 4 \Theta^{+}_1 - 4 \Phi_1^{+} + \sqrt{2} \Phi_3^{+} - \sqrt{2} \Theta_3^{+} )/3 } \\
                &\psi_{- \uparrow}(x,\tau) \sim \e^{-\iu ( 4 \Theta^{+}_1 + 4 \Phi_1^{+} - \sqrt{2} \Phi_3^{+} - \sqrt{2} \Theta_3^{+} )/3 } \\
                &\psi_{- \downarrow}(x,\tau) \sim \e^{-\iu \sqrt{2} ( \Theta^{+}_{2} + \Phi_{2}^{+} )}.
            \end{align}
    \end{subequations}
    Here we observe that the electron operators with positive helicity, $(+,\uparrow)$ and $(-,\downarrow)$ do not depend on the disordered field $\Theta^{+}_{1}$, while those with negative helicity, $(+,\downarrow)$ and $(-,\uparrow)$ do.
    As a result, the Green's functions of the negative-helicity electrons are exponentially decaying. 
    The electrons with positive helicity remain gapless and admit a description as a helical LL.
    For negative chirality of the spin chain $\chi=-$ the situation is reversed and the Green's functions of the positive-helicity electrons are exponentially decaying.
    This is the manifestation of the odd-parity spin splitting of the electrons in the odd-parity Kondo lattice.
    Depending on the chirality of the spin chain, one helical pair of electron modes is gapped out while the other remains gapless.
    To illustrate this result, we compute the spectral function of the gapless modes explicitly, following Ref.~\cite{Orignac-Suzumura-SpectralFunctionsTwoband-2011}.
    Writing 
    \begin{align}\label{eq:spectralfunction_I}
        A_{r\lambda}(q,\omega) = I_{r\lambda}(q,\omega)+I_{r\lambda}(-q,-\omega),
    \end{align}
    with 
    \begin{align}\label{eq:IfromG}
    \begin{split}
        I_{r\lambda}(q,\omega) &= \frac{1}{2\pi}\int \D t\,\D x\,
        \e^{-\iu(kx-\omega t)} G_{r\lambda}(x,t)
    \end{split}
    \end{align}
    we find for $(r,\lambda)$ such that $\eta = -\chi$ (see appendix \ref{app:spectralfunction_details}) 
    \begin{widetext}
        \begin{align}
            \begin{split}\label{eq:Irlambda_integral}
                I_{r\lambda}(q,\omega) &= 
                    \frac{\alpha^{\nu_1+\nu_2+\nu_1'+\nu_2'-1}}{\Gamma(\nu_1)\Gamma(\nu_2)\Gamma(\nu_1')\Gamma(\nu_2')}
                \int_0^1 \D w_1 \int_0^1 \D w_2 \, 
                w_1^{\nu_2-1}
                (1-w_1)^{\nu_1-1}     
                w_2^{\nu_2'-1}
                (1-w_2)^{\nu_1'-1} \\
                &\hspace{4em}\times 
                \frac{(\omega+ru(w_2)q)^{\nu_1+\nu_2-1}}{(u(w_1)+u(w_2))^{\nu_1+\nu_2-1}} \e^{-\alpha\frac{\omega+ru(w_2)q}{u(w_1)+u(w_2)}}
                \frac{
                (\omega-ru(w_1)q)^{\nu_1'+\nu_2'-1}}{(u(w_1)+u(w_2))^{\nu_1'+\nu_2'-1}}
                \e^{-\alpha\frac{\omega-ru(w_1)q}{u(w_1)+u(w_2)}}
                \\
                &\hspace{4em}
                \times
                \theta(\omega+r u(w_2)q)\theta(\omega-r u(w_1)q) 
                \phantom{\frac{1}{1}}
                ,
                \end{split}
        \end{align}
    \end{widetext}
    with 
    \begin{align}\label{eq:uofw}
        u(w)=u_2 w+u_1(1-w),
    \end{align}
    and
    \begin{subequations}\label{eq:exponents-spectralfunction}
        \begin{align}
        \nu_m &= \frac{1}{2}\left(\mathsf{M}_{1m} + (\mathsf{M}^{-1})_{m1}\right)^2 \\
        \nu_m' &= \frac{1}{2}\left(\mathsf{M}_{1m} - (\mathsf{M}^{-1})_{m1}\right)^2,
    \end{align}
    \end{subequations}
    for $m=1,2$
    where $\mathsf{M}$ is the $2\times2$ matrix which diagonalizes the Hamiltonian for $\Phi_2^\chi$ and $\Phi_3^\chi$ and $u_1<u_2$ are the corresponding velocities of the diagonal modes. 
    The integral in Eq.~\eqref{eq:Irlambda_integral} is rewritten in terms of hypergeometric functions as in Eqs.~(4.37-4.40) of Ref.~\cite{Orignac-Suzumura-SpectralFunctionsTwoband-2011} and evaluated as shown in Fig.~\ref{fig:spectral_function}.
    The dominant contribution of spectral weight comes from the power-law singularity at the edge $\omega=ru_1q$
    \begin{align}
        A_{r\lambda}(q,\omega\rightarrow ru_1q  + 0^+)\sim (\omega-ru_1q)^{\nu_2+\nu_1'+\nu_2'-1},
    \end{align}
    characteristic of LLs \cite{Meden-Schonhammer-SpectralFunctionsTomonagaLuttinger-1992,Voit-Voit-ChargespinSeparationSpectral-1993}.

    \section{Discussion and Conclusions}\label{sec:conclusion}

    In this paper we have explored a correlated model for an odd-parity magnet. 
    Motivated by the $\mathcal{T} \tau$ symmetry requirement for the magnetic state we wrote down a Hamiltonian for the odd-parity magnetic spin chain, from which we obtained a bosonic long-wavelength field theory via the Jordan-Wigner transformation and Abelian bosonization.
    We characterized the dominant correlations of the spin chain to make contact with the anticipated classical ground state.
    In particular, we found the spin chain to be chirally ordered owing to the relevance of the Dzyaloshinskii-Moriya interaction (DMI). 

    We extended the theory by including a separate chain of itinerant electrons Kondo-coupled to the local moments of the spin chain. 
    In the weak-coupling limit we obtained a bosonic long-wavelength field theory for the total system by bosonizing each chain separately. 
    At incommensurate fillings the low-energy electronic degrees of freedom retain a description in terms of a Luttinger liquid (LL) while the spin chain develops long-range chiral order. 
    Since the chiral spin chain contributes a gapless mode to the spectrum the total system resembles the fractional LL (LL$^*$) phase of the Kondo-Heisenberg lattice (KHL) \cite{Zhang-Vishwanath-PairdensitywaveSuperconductorDoping-2022,Tsvelik-Tsvelik-FractionalizedFermiLiquid-2016}, with two important distinctions.
    First, the spin chain is chirally ordered.
    This gives rise to a weak-coupling phase diagram that resembles that of the extended Hubbard model, with phase boundaries shifted by the Kondo coupling and a small region of dominant pair density-wave correlations. 
    The chiral order of the spin chain promotes composite order parameters to have power-law decay competing with purely electronic correlations.
    Second, the gapless mode of the spin chain is protected by the flow to strong coupling of the DMI, rather than the irrelevance of the forward-scattering term for ferromagnetic Kondo coupling. 
    This makes the present LL$^{*}$-like phase insensitive to the sign of $J_K$.
    
    When the electron filling is commensurate with the ordering wave vector of the spin chain $2k_F = |Q|$, back scattering of the electrons by the local moments becomes relevant, driving the LL$^{*}$-like phase to a phase with composite spin density-wave order, coexisting with the chiral order of the spin chain. 
    Importantly, the back scattering selects only electrons with helicity opposite of the chirality of the spin chain to enter in the SDW state, binding them in a spin singlet with the local moments, thus establishing a helical LL.
    Since the helicity is the product of spin and momentum, this mechanism furnishes the odd-parity spin polarization of the single-electron spectrum.
    This phase is more reminiscent of the $J_K > 0 $ KHL, which is characterized by a composite pair-density wave order \cite{Zachar-Tsvelik-OnedimensionalElectronGas-2001,Berg-Kivelson-PairDensityWaveCorrelationsKondoHeisenberg-2010}.
    Whereas the composite order in the KHL is driven by the Kondo coupling itself \cite{Zhang-Vishwanath-PairdensitywaveSuperconductorDoping-2022}, here it is driven by the commensuration of the electrons with the chiral spin background established by the DMI.
    
    The singlet binding is reminiscent of an orbital-selective Mott transition, in which one orbital of a multiorbital system localizes and becomes insulating \cite{Vojta-Vojta-OrbitalSelectiveMottTransitions-2010}. 
    Here it is not a given orbital but a specific helicity that is localized and the transition is driven by the interplay of commensuration and interactions rather than interactions alone.
    A similar instability toward a helical LL was found in Ref.~\cite{Tsvelik-Yevtushenko-PhysicsArbitrarilyDoped-2019}, where a ``helical metal" was realized at generic incommensurate fillings of the Kondo lattice. 
    In contrast to the present work, the authors treat the spin chain by imposing classical magnetic order with ordering wave vector $|Q| = 2k_F$ and consider fluctuations described by a nonlinear sigma model. 
    In the present work the fluctuations of the spin chain and the itinerant electrons are treated on equal footing in the Abelian bosonization framework, and the chiral order is established by the DMI.
    The strong spin-momentum locking of the spectral weight in the helical LL state parallels that of strongly correlated $d$-wave altermagnets \cite{Daghofer-vandenBrink-AltermagneticPolaronsFate-2026}, where one spin species is rendered incoherent by magnon scattering.
    The present mechanism of helicity-selective back scattering accompanied by a local moment spin-flip is similar, though here a genuine spectral gap is produced.
    
    The helicity-selective back scattering and the ensuing helical LL at commensurate filling is evidently the manifestation of odd-parity magnetism in this strongly correlated one-dimensional system. 
    Away from commensurate filling this interaction becomes irrelevant and the low-energy physics is governed by the LL$^{*}$-like phase, in which all remnants of odd-parity spin splitting are lost.
    To connect more naturally to material candidates for $p$-wave magnets, an extension of the present analysis to higher dimensions is of great interest.
    Such higher-dimensional generalizations can be constructed as arrays of weakly-coupled chains of the type studied in the present paper.
    This would add analytical insight into the ongoing debate around the heavy-fermion compound CeNiAsO, for which there has been reported conflicting evidence both for \cite{Zhang-Chen-OddspinsymmetryAnisotropy-2026} and against \cite{Zhang-Shen-QuenchingNonrelativistic$p$Wave-2026, Zhang-Zhou-SuppressionPwaveAltermagnetism-2026} the existence of $p$-wave spin splitting.
    This would also add a new layer of rigor to the construction of the two dimensional model of Ref.~\cite{Brekke-Linder-MinimalModelsTransport-2024}.
    Although our work investigates the origin of the electronic spin splitting, it does not address the mechanism giving rise to the chiral spin order.
    Whereas the DMI certainly furnishes the required chiral state, what remains to be understood is if and how such order can emerge from electronic correlations alone, not relying on spin-orbit coupling \cite{Kiselev-Schmalian-LimitsDynamicallyGenerated-2017,Wu-Chubukov-Conditions$l1$Pomeranchuk-2018}.
    This would clarify whether the spin splitting of the $p$-wave magnet can be considered a nonrelativistic phenomenon on the same footing as that of its cousin, the $d$-wave altermagnet. 

	\begin{acknowledgments}
        We thank Even Thingstad, Pavlo Sukhachov and Bjørnulf Brekke for helpful discussions. 
		We acknowledge support from the Norwegian Research Council through Grant No. 262633, ``Center of Excellence on Quantum Spintronics” and Grant No. 323766, as well as COST Action CA21144 ``Superconducting Nanodevices and Quantum Materials for Coherent Manipulation".
	\end{acknowledgments}

    \appendix

    \section{Details on the bosonization of the DMI term}\label{app:bosonization_details}
    Here we present the derivation of Eq.~\eqref{eq:H_D_bosonized} from Eq.~\eqref{eq:H_D_fermionized}.
    The odd and even sums are treated separately using 
    \begin{align}
        \begin{split}
            \e^{\iu 4k_F^Sx_j} = 
            \begin{cases}
                -1 & \mathrm{for}\, \ell=1       \\     
                +1 & \mathrm{for}\, \ell=2.
            \end{cases}
        \end{split}
    \end{align}
    Inserting Eq.~\eqref{eq:JW_fermions_left_right} and using the bosonized expression for the JW string operators on each chain $\ell=1,2$ \cite{Giamarchi-Giamarchi-QuantumPhysicsOne-2004}
    \begin{align}
        W_{j,\ell} = \frac{1}{2}(\e^{\iu k_F^S x_j - \iu \phi_\ell(x_j)} + \hc ),
    \end{align}
    we discard the oscillatory terms $\propto \e^{\pm\iu 2k_F^S x_j}$ and obtain
    \begin{align}
        \begin{split}
            &\frac{D}{2\iu} \sum_{j \, \text{odd}} \left[ d_{j,1}^{\dagger} W_{j,1}^{\mathstrut} W_{j+1,2}^{\mathstrut} d_{j+1,2}^{\mathstrut} - \hc  \right]   \\
            &=\frac{D 2a}{4\cdot2\pi\alpha} 
            \sum_{j \,\mathrm{odd}}
            \Bigl[4\sin(\theta_2(x_{j+1})-\theta_1(x_j) ) \\
            &+\cos(2\phi_1(x_j)-2\phi_2(x_{j+1}) - \theta_1(x_j) + \theta_2(x_{j+1})) \phantom{\frac{1}{1}}  \\
            &-\cos(-2\phi_1(x_j)+2\phi_2(x_{j+1}) - \theta_1(x_j) + \theta_2(x_{j+1})) \phantom{\frac{1}{1}}  \\
            &+\cos(2\phi_1(x_j)+2\phi_2(x_{j+1}) - \theta_1(x_j) + \theta_2(x_{j+1})) \phantom{\frac{1}{1}}  \\
            &-\cos(-2\phi_1(x_j)-2\phi_2(x_{j+1}) - \theta_1(x_j) + \theta_2(x_{j+1}))
            \Bigr] ,\phantom{\frac{1}{1}} 
        \end{split}
    \end{align}
    after resolving $\e^{\pm\iu 2k_F^S a} = \pm\iu$. 
    Performing an analogous computation for the even sum in Eq.~\eqref{eq:H_D_fermionized} and adding up the contributions one finds that in the continuum limit $\phi_\ell(x_j)\approx\phi_\ell(x_{j+1})\rightarrow\phi_\ell(x)$, the cosine terms cancel while the sine terms add up yielding Eq.~\eqref{eq:H_D_bosonized}.

    \onecolumngrid

    \section{Details on the computation of the spectral function}\label{app:spectralfunction_details}
    Here we summarize the main details of the calculation of the spectral function for a generic two-component LL following Ref.~\cite{Orignac-Suzumura-SpectralFunctionsTwoband-2011}.
    The Green's function for the gapless electrons with $\eta=\chi$ is obtained as 
    \begin{align}
    \begin{split}
        G_{r\lambda}(x,t) &= \frac{\e^{-\iu r k_Fx}}{2\pi\alpha}\prod_{m=1,2} \left(\frac{\alpha}{\alpha+\iu(u_mt-rx)}\right)^{\nu_m} 
        \left(\frac{\alpha}{\alpha+\iu(u_mt+rx)}\right)^{\nu_m'},
    \end{split}
    \end{align}
    where the exponents are given by Eq.~\eqref{eq:exponents-spectralfunction}.
    Using $\overline{G_{r\lambda}(x,t)} = G_{r\lambda}(-x,-t)$ one can write the spectral function as in Eq.~\eqref{eq:spectralfunction_I} with $I_{r\lambda}(q,\omega)$ as in Eq.~\eqref{eq:IfromG}.
    Using the Feynman identity
    \begin{align}
        \frac{1}{X_1^{\gamma_1} X_2^{\gamma_2}} = \frac{\Gamma(\gamma_1+\gamma_2)}{\Gamma(\gamma_1)\Gamma(\gamma_2)} \int_0^1 \D w \frac{w^{\gamma_1-1}(1-w)^{\gamma_2-1}}{\left[X_1w+X_2(1-w)\right]^{\gamma_1+\gamma_2}},
    \end{align}
    valid for $\gamma_1,\gamma_2>0$, we find
    \begin{align}
        \begin{split}\label{eq:Irlambda-preFourier}
            I_{r\lambda}(q,\omega) &= \frac{\alpha^{\nu_1+\nu_2+\nu_1'+\nu_2'-1}}{(2\pi)^2}  \int \D t \, \D x \, \e^{-\iu(qx-\omega t)} \frac{\Gamma(\nu_1+\nu_2)\Gamma(\nu_1'+\nu_2')}{\Gamma(\nu_1)\Gamma(\nu_2)\Gamma(\nu_1')\Gamma(\nu_2')} 
            \\
            &\hspace{1em}\times \int_0^1 \D w_1 \int_0^1 \D w_2
            \frac{w_1^{\nu_2-1}(1-w_1)^{\nu_1-1}}{\left[\alpha + \iu( u(w_1)t-rx)\right]^{\nu_1+\nu_2}}
            \frac{w_2^{\nu_2'-1}(1-w_2)^{\nu_1'-1}}{\left[\alpha + \iu( u(w_2)t+rx)\right]^{\nu_1'+\nu_2'}},
        \end{split}
    \end{align}
	with $u(w)$ given by Eq.~\eqref{eq:uofw}.
    By changing variables to 
    \begin{align}
        \xi_1 = u(w_1)t-rx \qquad \text{and} \qquad
        \xi_2 = u(w_2)t+rx,
    \end{align}
    and using the integral identity 
    \begin{align}
        \int_{-\infty}^\infty \D x \, (b+\iu x)^{-\nu} \e^{-\iu px}
        =
        \theta(-p)\frac{2\pi (-p)^{\nu-1} \e^{bp}}{\Gamma(\nu)} \quad \text{for} \quad \Re b>0,\Re \nu >0,
    \end{align}
    from Eq.~(3.382.6) in Ref.~\cite{Gradshteyn-Ryzhik-TableIntegralsSeries-2014} one can perform the Fourier integration in Eq.~\eqref{eq:Irlambda-preFourier} to arrive at Eq.~\eqref{eq:Irlambda_integral}.
    
    \twocolumngrid

    \bibliography{references}
    
\end{document}